**TITLE**

Mid-infrared Spectroscopic Observations of the Dust-forming Classical Nova V2676 Oph [*]

**AUTHORS**

Hideyo Kawakita[1,2]; e-mail: kawakthd@cc.kyoto-su.ac.jp

Takafumi Ootsubo[3],

Akira Arai[1]

Yoshiharu Shinnaka[1,2,4]

&

Masayoshi Nagashima[2]

**AFFILIATIONS**

[1]Koyama Astronomical Observatory, Kyoto Sangyo University, Motoyama, Kamigamo, Kita-ku, Kyoto 603-8555, Japan

[2]Department of Physics, Faculty of Science, Kyoto Sangyo University, Motoyama, Kamigamo, Kita-ku, Kyoto 603-8555, Japan

[3]Department of Earth Science and Astronomy, Graduate School of Arts and Sciences, The University of Tokyo, 3-8-1 Komaba, Meguro-ku, Tokyo 153-8902, Japan

[4]National Astronomical Observatory of Japan, 2-21-1 Osawa, Mitaka, Tokyo 181-8588

**ABSTRACT**

The dust-forming nova V2676 Oph is unique in that it was the first nova to provide evidence of $C_2$ and CN molecules during its near-maximum phase and evidence of CO molecules during its early decline phase. Observations of this nova have revealed the slow evolution of its lightcurves and have also shown low isotopic ratios of carbon ($^{12}C/^{13}C$) and nitrogen ($^{14}N/^{15}N$) in its nova envelope. These behaviors indicate that the white dwarf (WD) star hosting V2676 Oph is a CO-rich WD rather than an ONe-rich WD (typically larger in mass than the former). We performed mid-infrared spectroscopic and photometric observations of V2676 Oph in 2013 and 2014 (respectively 452 and 782 days after its discovery). No significant [Ne II] emission at 12.8 µm was detected at either epoch. These provided evidence for a CO-rich WD star hosting V2676 Oph. Both carbon-rich and oxygen-rich grains were detected in addition to an unidentified infrared feature at 11.4 µm originating from polycyclic aromatic hydrocarbon molecules or hydrogenated amorphous carbon grains in the envelope of V2676 Oph.

**1. Introduction**

A classical nova is an explosive event that occurs on a white dwarf (WD) star, which causes it to suddenly brighten. Its development begins with two main stars in a closed binary system; one star evolves into a red giant leaving its remnant WD core in orbit with the remaining star. The second star sheds its envelope into the WD companion when its Roche lobe has filled. The explosion of the gas envelope of the WD (accreted from the companion star) is powered by thermonuclear runaway (TNR) reactions that occur at the base of the envelope (José $et\ al.$ 2006). The strength of the explosion depends on multiple physical parameters of the WD and accreted mass from the companion star. The key parameter is the amount of pressure achieved at the core-envelope interface (proportional to the WD mass), with the process of mass ejection requiring pressure in the range of $\sim 10^{19}$–$10^{20}$ dyn cm$^{-2}$, depending on the chemical composition of the gas as a mixture of materials accreted from both the companion star and the WD surface (Shara 1981; Fujimoto 1982; MacDonald 1983).

The chemistry of the nucleosynthesis products from the TNR reactions depends on the thermal history of the explosion, in particular, the peak temperatures achieved at the base of the envelope, the characteristic timescales of the TNR reactions, and the initial elemental abundances of the gas mixture. The nature of the WD star (i.e., whether it is CO-rich or ONe-rich) is essential to the chemistry of the ejected material in novae. The typical mass of a CO-rich WD is smaller than that of an ONe-rich WD, and their masses are approximately distinguished by $\sim 1.1$ M$_{sun}$ when binarity is taken into account (Gil-Pons $et\ al.$ 2003). The base of the envelope on ONe-rich WD stars tends to have higher peak temperatures than that of CO-rich WDs during TNR reactions (José & Hernanz 1998) due to the greater degenerate conditions in more massive WDs (Starrfield 2008). In addition, the chemical composition of the gas mixture between the accreted (solar-type) gas and surface of both types of WDs is different; thus, they produce ejectae from the TNR reactions with very different chemical compositions.

The classical nova V2676 Oph, which is the focus of this paper, is unique because it was the first nova to provide evidence of $C_2$ molecules, and is only the second example of a nova in which CN molecules have been detected (the first example was DQ Her in 1934) during its near-maximum phase, as reported by Nagashima $et\ al.$ (2014). The nova was discovered by Nishimura (2012) with a UT on 2012 March 25.8 (probably just after the explosion) and was considered an "Fe II-type" classical nova (according to the classification by Williams 1992) based on optical low-resolution spectroscopic observations (Arai & Isogai 2012). Kawakita $et\ al.$ (2016) revealed that the photospheric temperature of V2676 Oph had cooled to $\sim 4500$ K around its visual brightness maximum; $C_2$ and CN molecules form while the estimated typical photospheric temperature of a nova is $\sim 8000$ K (Evans $et\ al.$ 2005), corresponding to F-type supergiant stars. During the early decline phase of the nova, strong emission from CO molecules was also detected (Rudy $et\ al.$ 2012). The existence of both $C_2$ and CN radicals in the photosphere indicates that the nova envelope was C-rich, with C/O>1 (Kawakita $et\ al.$ 2016).

To understand the peculiarity of V2676 Oph, we conducted mid-infrared ($N$-band) spectroscopic observations of the nova during its late decline phase. In particular, the [Ne II] emission line at 12.8 μm in the nebular phase of a nova provides a clue to the nature of the WD star hosting V2676 Oph with regard to whether it is CO- or ONe-rich. Usually, novae occurring on ONe-rich WDs exhibit strong [Ne II] emission lines at 12.8 μm in the $N$-band (Gehrz 2008; Evans & Gehrz 2012). Furthermore, V2676 Oph exhibits the signature of dust formation in its light curves (Raj $et\ al.$ 2016). Understanding the nature of dust grains formed in V2676 Oph is also interesting from the point of view of the dust-formation process in novae. Because the envelope of the nova is carbon-rich, C/O>1 (i.e., oxygen atoms are consumed by the formation of CO molecules if the chemical equilibrium for molecular and dust formation is assumed), carbon grains rather than silicate grains in V2676 Oph are, in principle, expected to be enriched. However, multiple types of grains have been simultaneously observed in some novae (Gehrz 2008; Evans & Gehrz 2012). Typically, carbon-rich grains such as amorphous carbon grains form earlier in the outburst, and oxygen-rich grains such as silicate grains form later (Gehrz *et al*. 1998; Sakon *et al*. 2016). The coexistence of thermal emission from carbon-rich dust grains with unidentified infrared (UIR) features, usually attributed to polycyclic aromatic hydrocarbon (PAH) molecules or hydrogenated amorphous carbon (HAC) grains, is also interesting from the viewpoint of the origin of UIR carriers in novae (Evans & Rawlings 2008). PAH molecules probably originate from simple diatomic molecules such as $C_2$ via chemical reactions, and the UIR feature may be caused by free-flying PAH molecules or PAH-like molecules loosely aggregated into carbon particles such as HAC grains (Evans *et al*. 1997).

**2. Observations and Data Reduction**

Imaging and low-resolution spectroscopic observations of the classical nova V2676 Oph in the *N*-band (8–13 µm) were performed with the Cooled Mid-Infrared Camera and Spectrometer (COMICS) mounted on the 8 m Subaru telescope in Hawaii (Kataza *et al.* 2000; Okamoto *et al.* 2003; Sako *et al.* 2003) on UT 2013 June 20.4 and UT 2014 May 16.5 (452 days and 782 days respectively after the discovery of the nova). We used a slit of 0.33 arcsec in width and 40 arcsec in length for the low-resolution mode of the *N*-band, corresponding to a spectral resolving power of $R\sim250$. Imaging photometry observations with narrow-band filters centered at 7.8, 8.8, and 12.4 µm (the standard filters installed on COMICS) were also performed to precisely calibrate the flux of the observed spectra. In addition to the nova, HD 169916 (for both photometry and spectroscopy in 2013), HD 186791, and HD 152880 (for photometry and for spectroscopy in 2014) were also observed as standard stars (Cohen *et al.* 1999) for complete flux calibration and correction of atmospheric absorption. The observational circumstances are listed in Table 1.

To cancel the high background radiation, secondary mirror chopping was used at a frequency of ~0.45 Hz. Chopping throws in the north-south direction had a width of 15 arcsec for the observation of the nova in 2014, and 10 arcsec for other observations. In 2014, we employed nodding for the imaging observation of the nova because it was faint, whereas other observations were conducted without nodding. For the spectroscopy, the position angle of the slit was set to 0 degrees so that images at both chopping positions would fall on it. Data reduction was followed by the standard reduction procedure provided by the COMICS instrument team
(http://www.naoj.org/Observing/DataReduction/Cookbooks/COMICS_COOKBOOK.pdf).
To improve the signal-to-noise ratio of the data, we used images and spectra collected at both chopping positions for the data reduction. The co-added on-source integration time for each object is listed in Table 1. The aperture size for the photometry was set to 1.3 arcsec in radius. The spectra were extracted with a slit region of 0.33 arcsec (slit width) by 1.49 and 1.16 arcsec (along the slit direction) for data taken in 2013 and 2014, respectively, to maximize the signal-to-noise ratios of the extracted spectra. Slit-loss corrections for the spectra were adopted based on the photometric data.

Table 1. Observational logs.

| UT Date | Target | Integration Time [s] | Obs.-mode | Air Mass |
|---|---|---|---|---|
| 2013 Jun 20.41 | V2676 Oph | 240 | Spectroscopy (NL-mode) | 1.431 |
| 2013 Jun 20.41 | V2676 Oph | 60 | Photometry (12.4 μm) | 1.430 |
| 2013 Jun 20.40 | V2676 Oph | 20 | Photometry (8.8 μm) | 1.436 |
| 2013 Jun 20.42 | V2676 Oph | 120 | Photometry (7.8 μm) | 1.432 |
| 2013 Jun 20.42 | HD 169916 | 20 | Spectroscopy (NL-mode) | 1.446 |
| 2013 Jun 20.43 | HD 169916 | 10 | Photometry (12.4 μm) | 1.439 |
| 2013 Jun 20.42 | HD 169916 | 10 | Photometry (8.8 μm) | 1.441 |
| 2013 Jun 20.42 | HD 169916 | 10 | Photometry (7.8 μm) | 1.443 |
| 2014 May 16.54 | V2676 Oph | 480 | Spectroscopy (NL-mode) | 1.465—1.484 |
| 2014 May 16.51 | V2676 Oph | 400 | Photometry (12.4 μm) | 1.432 |
| 2014 May 16.50 | V2676 Oph | 400 | Photometry (8.8 μm) | 1.431 |
| 2014 May 16.49 | V2676 Oph | 400 | Photometry (7.8 μm) | 1.433 |
| 2014 May 16.55 | HD 152880 | 20 | Spectroscopy (NL-mode) | 1.370 |
| 2014 May 16.48 | HD 186791 | 10 | Photometry (12.4 μm) | 1.336 |
| 2014 May 16.49 | HD 186791 | 10 | Photometry (8.8 μm) | 1.325 |
| 2014 May 16.49 | HD 186791 | 10 | Photometry (7.8 μm) | 1.313 |

**3. Results and Discussion**

Figure 1 shows the observed low-resolution spectra of V2676 Oph in the *N*-band at $t$ = 452 day and 782 day ($t$ = 0 day is taken to mark the discovery). No clear emission of [Ne II] at 12.8 μm was seen in either the 2013 or 2014 observations, although the [Ne II] emission line is usually recognized in the nebula phase of a nova originating from an ONe-rich WD. To confirm the presence of the [Ne II] emission line, we subtracted the synthesized thermal emission of grains from the observed spectra in the manner described below. In subsequent paragraphs, we will discuss the presence of [Ne II] emission based on the residuals of the spectra. We will also discuss the nature of the dust grains formed in V2676 Oph, as well as the UIR features.

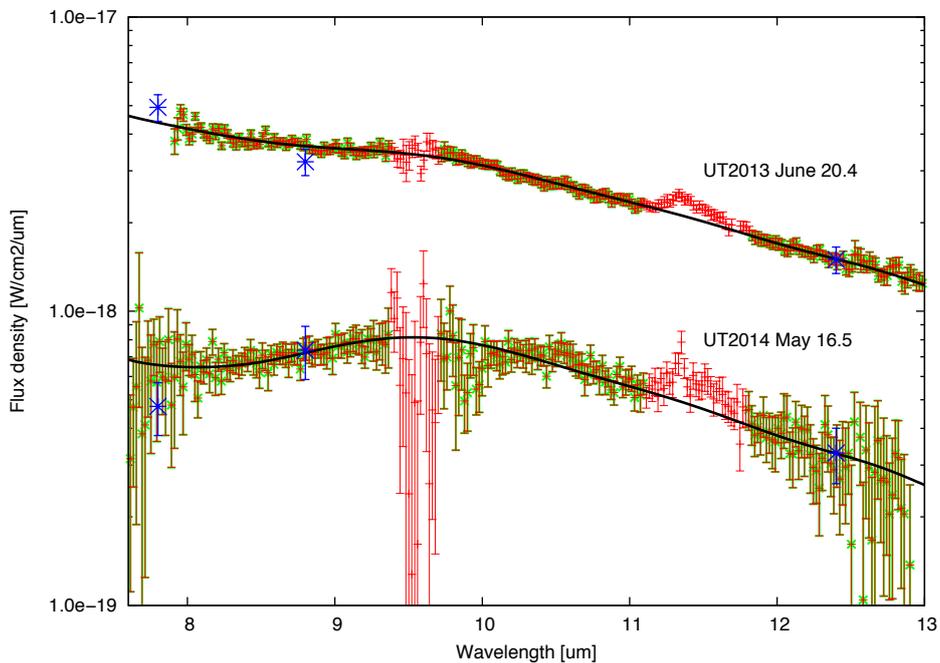

Figure 1: Low-resolution *N*-band spectra of V2676 Oph taken at $t$ = 452 day (UT 2013 June 20.4) and 782 day (UT 2014 May 16.5). Photometric data points on both dates are also plotted in the same figure. Synthesized dust emission spectra (thick line) are overplotted for comparison. To fit the synthesized dust emission spectrum to the observed spectrum at each epoch, we eliminated that part of the observed data (shown with thin error bars) corresponding to the wavelength ranges of both the telluric ozone absorption band at ~9.6 μm and the UIR emission at 11.4 μm.

3.1 Dust Grains

To reproduce the observed thermal emission from dust grains, we assumed two types of grains: amorphous carbon and astronomical silicate. The optical constants of these types of grain are referenced in Preibisch *et al.* (1993) and Draine (1985), respectively. The thermal emission from dust grains at λ (μm) is modeled by the following expression in the case of the optically thin condition:

$$f_{\text{NOVA}}(\lambda) = q \times [\,(1-r) \times Q_{\text{abs}}[\text{aCar}] \times \text{BB}(T_{\text{aCar}}) + r \times Q_{\text{abs}}[\text{astrSil}] \times \text{BB}(T_{\text{astrSil}})\,]$$
$$\times f_{\text{ext}}(\lambda)\,.$$

The parameter $q$ denotes the scaling factor for the observed flux density in [W m$^{-2}$ μm$^{-1}$], $r$ is the fraction of astronomical silicate grains, and $Q_{abs}$[aCar] and $Q_{abs}$[astrSil] are the absorption efficiencies of amorphous carbon and astronomical silicate grains, respectively. BB($T$) is the Planck function at temperature $T$. Interstellar extinction, $f_{ext}(\lambda)$, is modeled in a similar way to Sakon *et al*. (2016). We assumed astronomical silicate grains of a 0.1 μm radius to reproduce the shape of the extinction curve in the *N*-band. The optical depth at 9.7 μm for interstellar extinction ($\tau_{9.7}$) was calibrated by the estimated value of $\tau_{9.7}$ based on extinction in the *V*-band ($A_V$) towards the direction of V2676 Oph, $A_V = 2.65 \pm 0.15$ (Kawakita *et al*. 2016). We converted the value of $A_V$ to $\tau_{9.7}$ based on the formula: $A_V/\tau_{9.7} = 18.5 \pm 1.5$ (Roche & Aitken 1984). Finally, we fitted the parameters to reproduce the observed spectra using the least-$\chi^2$ method. We eliminated the part of the observed data corresponding to the wavelength ranges for both a telluric ozone absorption band (~9.6 μm) and UIR emission at ~11.4 μm.

Synthesized spectra for the observations are also shown in Figure 1. Table 2 lists the parameters that fit with the observed spectra. The masses of the carbon and silicate grains can be derived from the parameters $q$ and $r$ if the radius and density of each type of grain are assumed. However, the radii of dust grains are not well constrained in our case. Our observations are only in the mid-infrared wavelength region from 8 to 13 μm, and the plausible sizes of the grains (smaller than ~1 μm, as in the cases of V705 Cas (Evans *et al*. 2005) and V1280 Sco (Sakon *et al*. 2016)) correspond to the Rayleigh regime for the absorbing properties of the dust. Therefore, the shapes of the dust emission spectra are similar for dust grains smaller than ~0.5 μm, and it is difficult to constrain a lower limit for the dust radii. We assumed the grain radii for amorphous carbon and astronomical silicate to be 0.1 μm (if grain radii larger than ~1 μm are used, the $\chi^2$ values become much worse). To provide stronger constraints for the dust radii we would need a greater number of measurements in shorter and longer wavelength regions at each epoch. Even if we changed the grain radii (to smaller than ~0.5 μm), the derived dust temperatures and the least-$\chi^2$ values would not change dramatically, as shown in Tables 3 and 4. Instead, we will use the parameters listed in Table 2 for further discussion.

Table 2. Best-fit parameters for dust emission $^{(*)}$.

| Epoch (t = 0 day at Discovery) | $q$ | $r$ | $T_{aCar}$ (K) | $T_{astrSil} / T_{aCar}$ | Reduced-$\chi^2$ |
|---|---|---|---|---|---|
| $t$ = 452 day | 5.5±0.7 | 0.83±0.05 | 530±30 | 0.45±0.04 | 0.71 |
| $t$ = 782 day | 1.3±0.2 | 0.98±0.03 | 840±460 | 0.30±0.17 | 0.72 |

$^{(*)}$ The grain radii for amorphous carbon and astronomical silicate are assumed to be 0.1 μm.

Table 3. Best-fit parameters for dust emission $^{(*)}$.

| Epoch (t = 0 day at Discovery) | $q$ | $r$ | $T_{aCar}$ (K) | $T_{astrSil} / T_{aCar}$ | Reduced-$\chi^2$ |
|---|---|---|---|---|---|
| $t$ = 452 day | 1.8±0.1 | 0.28±0.07 | 480±20 | 0.54±0.04 | 0.74 |
| $t$ = 782 day | 0.22±0.03 | 0.82±0.19 | 770±350 | 0.35±0.17 | 0.70 |

$^{(*)}$ The grain radii for amorphous carbon and astronomical silicate are assumed to be 0.1 and 0.5 μm, respectively.

Table 4. Best-fit parameters for dust emission ⁽*⁾.

| Epoch (t = 0 day at Discovery) | $q$ | $r$ | $T_{\mathrm{aCar}}$ (K) | $T_{\mathrm{astrSil}} / T_{\mathrm{aCar}}$ | Reduced-$\chi^2$ |
|---|---|---|---|---|---|
| $t$ = 452 day | 13±2 | 0.04±0.01 | 480±20 | 0.54±0.04 | 0.74 |
| $t$ = 782 day | 0.57±0.44 | 0.31±0.28 | 770±350 | 0.35±0.17 | 0.70 |

⁽*⁾ The grain radii for amorphous carbon and astronomical silicate are assumed to be 0.01 and 0.5 μm, respectively.

Although the dust mass cannot be well constrained, the presence of silicate grains in the carbon-rich ejecta of V2676 Oph is interesting from the point of view of grain formation in novae. In the carbon-rich (C/O>1) ejecta of V2676 Oph, we detected silicate (oxygen-rich) grains at both epochs, as well as carbon-rich grains. The presence of multiple types of grain was also confirmed in a number of other novae (Gehrz *et al*. 1998; Gehrz 2008; Evans & Gehrz 2012; Sakon *et al*. 2016) even in the carbon-rich envelopes of some novae such as V1370 Aql, V842 Cen (Andreä *et al*. 1994), and QV Vul (Gehrz *et al*. 1992).

The temperatures of the astronomical silicate grains were cooler than those of amorphous carbon grains by a factor of around 2–3. This might be caused by different refractive indices for different types of dust grain (higher absorption coefficients for amorphous carbon lead to higher temperatures than for astronomical silicate). Regarding the temporal evolution of dust temperatures, the dust temperature $T_d$ is expected to decline with time $t$ as $T_d \propto t^{-2/(\beta+4)}$, if we assume the constant luminosity of the central nova and unchanged dust properties between the epochs of our observations (Evans & Gehrz 2012). It is difficult to discuss the β-index of the dust grains (Evans *et al*. 1997) because of the large error in the dust temperature at $t$ = 782 day.

3.2 [Ne II] Emission at 12.8 μm

The residual spectra of V2676 Oph after the subtraction of synthesized dust emission spectra from the observed spectra (Fig. 2) showed no evidence of the strong emission of [Ne II] at 12.8 μm, which is usually observed in the late phase of a nova occurring on an ONe-rich WD (Gehrz *et al*. 1998; Gehrz 2008; Evans & Gehrz 2012). This suggests that V2676 Oph originated on a CO-rich WD. The dust formation in V2676 Oph also supports this hypothesis, because novae on ONe-rich WDs exhibit little dust formation during their decline phase (Gehrz 2008).

The lightcurve evolution of V2676 Oph is slower than that of typical novae (Nagashima *et al*. 2014), and it is believed that such slow novae occur on WDs that have relatively smaller masses (Hachisu & Kato 2006). The similarity between the lightcurves of V2676 Oph and DQ Her (in 1934) implies that the mass of the WD star hosting V2676 Oph was similar to that of DQ Her, ~0.6 $M_{\mathrm{sun}}$ (Horne *et al*. 1993; Hachisu & Kato 2015). Because the typical mass of CO-rich WDs is smaller than that of ONe-rich WDs, and their masses may be distinguished by ~1.1 $M_{\mathrm{sun}}$ (Gil-Pons *et al*. 2003), the WD mass estimated from the lightcurves of V2676 Oph is consistent with the absence of the [Ne II] emission in our spectra.

However, the presence of $C_2$ and CN molecules in V2676 Oph indicates a carbon-rich envelope (C/O>1), which is inconsistent with the calculations of TNR theory for CO-rich WDs (José & Hernanz 1998; Denissenkov *et al.* 2014). The C/O abundance ratio of a nova's ejecta is expected to be smaller than unity in the case of CO-rich WDs, and the carbon-rich envelope is only possible in the case of ONe-rich WDs based on the current TNR theory. However, Denissenkov *et al.* (2014) pointed out that the C/O ratio varies within the envelope shell, depending on the distance from the hosting WD star. Only some parts of the shell may have C/O>1. Furthermore, it is worth noting that nova models have traditionally assumed that the composition of C/O is ~1, as noted by José *et al.* (2016). Recent calculations for the interior structure of CO-rich WDs predict C/O>1 for the outer layer of the WD (Fields *et al.* 2016; José *et al.* 2016). Based on such a carbon-rich outer layer of WDs hosting novae, the carbon-rich ejecta of novae are able to be reproduced (José *et al.* 2016).

As regards the isotopic ratios found in V2676 Oph ($^{12}C/^{13}C$ ~4 and $^{14}N/^{15}N$ ~2; Kawakita *et al.* 2015), the calculations by José & Hernanz (2007) were not able to reproduce its observed isotopic ratios. On the other hand, recent calculations by Denissenkov *et al.* (2014) reproduced similar values to those observed in V2676 Oph in the cased of a CO-rich WD with a mass of ~1.15 $M_{sun}$ (Fig. 21 in Denissenkov *et al.* 2014). Such a large mass for a CO-rich WD may be inconsistent with the smaller mass estimated from the aforementioned lightcurves.

We also note that the accretion of solar-type gas from the companion to the WD is usually considered in the TNR calculation. The companion star in V2676 Oph might be a carbon-rich star that can feed gas enriched in carbon (and the isotopic ratios of carbon and nitrogen are quite different from the solar values) to the WD hosting V2676 Oph.

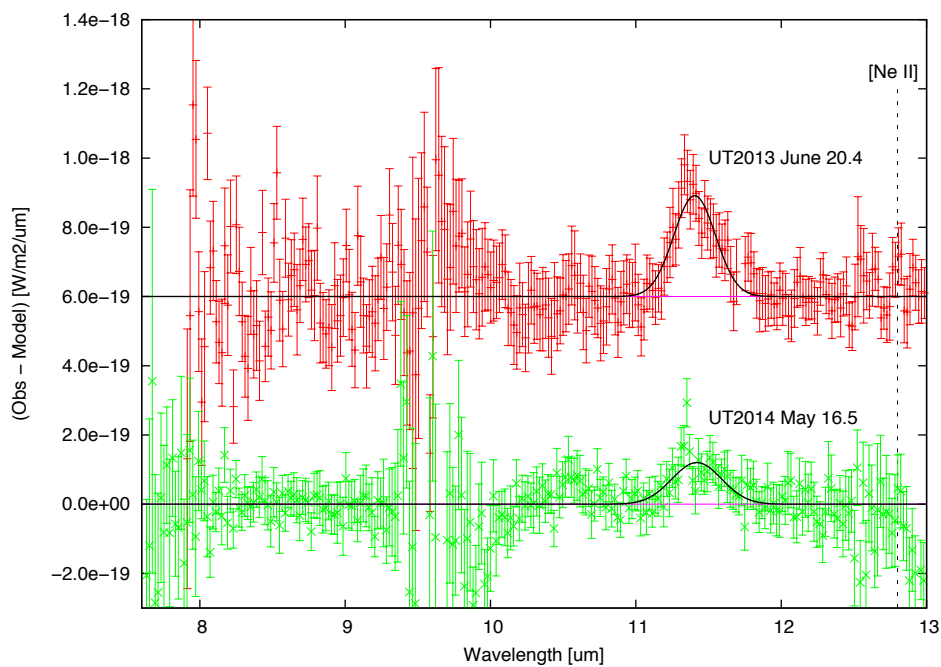

Figure 2: Residual spectra of V2676 Oph at $t$ = 452 day (UT 2013 June 20.4) and 782 day (UT 2014 May 16.5). The residual spectrum at $t$ = 452 day is shifted by $6\times10^{-19}$ to distinguish these spectra. No prominent [Ne II] emission line at 12.8 μm is recognized in the observed spectra (there is a hint of [Ne II] emission but not significantly so in 2013). The UIR emission

at ~11.4 µm is fitted with a Gaussian function. A broad emission around 9.7 µm at $t = 452$ day is considered a residual of the 9.7 µm silicate feature.

3.3 The UIR feature

The UIR emission at 11.4 µm (identified as the aromatic C-H "out-of-plane bending" feature) was clearly detected in both spectra taken at $t = 452$ day and 782 day. The coexistence of UIR carriers and carbon-rich grains in V2676 Oph supports the formation of UIR carriers from amorphous carbon grains, as predicted by Mitchell & Evans (1984). If we fit the UIR emission with a Gaussian function, the central wavelength (and full width at half maximum) are 11.40 ± 0.01 µm (0.34 ±0.02 µm) at $t = 452$ day and 11.42 ± 0.02 µm (0.40 ± 0.05 µm) at $t = 782$ day, and are consistent with those of the UIR emission in V705 Cas (Evans *et al.* 2005). These central wavelengths are slightly longer than the standard wavelength of the UIR feature, 11.25 µm (Evans & Gehrz 2012). No other strong UIR features are seen in V2676 Oph, although a prominent emission complex at around 8.0–9.0 µm was observed in V705 Cas, as reported by Evans *et al.* (2005). The absence of the expected UIR emission at ~8.6 µm (weaker than the emission at ~11.4 µm) might indicate that PAHs in V2676 Oph at $t = 452$ day and 782 day were neutral PAHs rather than ionized PAHs (see Fig. 3 in Draine & Lee 2007). However, because free-flying PAH molecules cannot survive against the strong radiation field from the central nova in the later phase, the observed UIR might be explained instead by HAC grains (Evans & Rawlings 1994). The weak emission feature at ~10.6 µm at $t = 782$ day cannot be assigned to a known UIR feature.

**4. Conclusion**

    Mid-infrared low-resolution spectroscopic and photometric observations of V2676 Oph were performed by COMICS mounted on the 8 m Subaru telescope on UT 2013 June 20 and UT 2014 May 16 (482 days and 782 days respectively after its discovery). Both types of dust grain, carbon-rich and oxygen-rich grains, were detected on both dates. The 11.4 μm UIR emission was also detected on these dates. However, no clear [Ne II] emission at 12.8 μm was detected. Based on the absence of [Ne II] emission, the WD hosting V2676 Oph is considered a CO-rich WD. This hypothesis is supported by dust production in the nova and the slow evolution of its lightcurves. However, the observed isotopic ratios of carbon and nitrogen in V2676 Oph can be reproduced by the nucleosynthesis models for WDs with a moderate mass (~1.1 $M_{sun}$) close to the high-end of the mass range for CO-rich WDs (Denissenkov *et al*. 2014). The carbon-rich chemistry of the ejecta (C/O>1) as well as the low isotopic ratios ($^{12}C/^{13}C$ and $^{14}N/^{15}N$) observed in V2676 Oph (Kawakita *et al*. 2015, 2016) challenge the traditional TNR models for low-mass CO-rich WDs (José & Hernanz 1998, 2007; José *et al*. 2006, 2016; Denissenkov *et al*. 2014).